\documentclass{article}

\usepackage[preprint]{neurips_2025}
\usepackage[utf8]{inputenc} 
\usepackage[T1]{fontenc}    
\usepackage{hyperref}       
\usepackage{url}            
\usepackage{booktabs}       
\usepackage{amsfonts}       
\usepackage{nicefrac}       
\usepackage{microtype}      
\usepackage{xcolor}         
\usepackage{amsmath}

\title{Regulating the Agency of LLM-based Agents}
  
\author{%
 Se\'an Boddy\\
  Berkman Klein Center for Internet \& Society \\
  Harvard University \\
  \texttt{sboddy@law.harvard.edu}
    \And
  Joshua Joseph\\
  Berkman Klein Center for Internet \& Society \\
  Harvard University \\
  \texttt{jjoseph@law.harvard.edu}
}

\begin{document}

\maketitle

\vspace{-0.5cm}

\begin{abstract}

As increasingly capable large language model (LLM)-based agents are developed, the potential harms caused by misalignment and loss of control grow correspondingly severe. To address these risks, we propose an approach that directly measures and controls the agency of these AI systems. We conceptualize the agency of LLM-based agents as a property independent of intelligence-related measures and consistent with the interdisciplinary literature on the concept of agency. We offer (1) agency as a system property operationalized along the dimensions of preference rigidity, independent operation, and goal persistence, (2) a representation engineering approach to the measurement and control of the agency of an LLM-based agent, and (3) regulatory tools enabled by this approach: mandated testing protocols, domain-specific agency limits, insurance frameworks that price risk based on agency, and agency ceilings to prevent societal-scale risks. We view our approach as a step toward reducing the risks that motivate the ``Scientist AI'' paradigm, while still capturing some of the benefits from limited agentic behavior.

\end{abstract}

\section{Introduction}

As large language model (LLM)-based agents become increasingly capable of performing complex, multi-step tasks in real-world environments, the need for effective control mechanisms becomes critical \citep{toner2024chat, bengio2025internationalscientificreportsafety}. Recent work has underscored this need by demonstrating behaviors such as deception, blackmail, goal-guarding, and resistance to shutdown attempts \citep{pan2024feedback, meinke2025frontier, lynch2025agentic}. These behaviors emerge as ordinary consequences of the current training paradigms, and these risks increase with the capabilities of the model \citep{dung2023current, bengio2025superintelligent}. As a result, misaligned agents have the potential to cause even greater harms by resisting human control, self-replicating, and disrupting critical infrastructure \citep{mitre2025artificial, clymer2024rogue}. 

Despite these risks, we lack tools and frameworks to measure or control the source of the problematic behavior: the system's agency \citep{bengio2025superintelligent}. Therefore, our aim is to make the agency of an LLM-based agent the direct target of regulatory observation and intervention. We do this by conceptualizing agency as a measurable system property distinct from intelligence and operationalizing it along the dimensions of preference rigidity, independent operation, and goal persistence, which are consistent with the interdisciplinary literature on the concept of agency. We propose a representation engineering \citep{zou2025representationengineeringtopdownapproach} approach, which builds on \citet{chen2024designingdashboardtransparencycontrol} and trains linear probes and hypothesize that these probes allow for both the measurement and control of agency. These ``agency sliders,'' function similarly to \citet{chen2024designingdashboardtransparencycontrol}'s dashboard sliders for the LLM's representation of the user's age, gender, educational level, and socioeconomic status (see Figure 2 from \citet{chen2024designingdashboardtransparencycontrol}). This approach enables a variety of regulatory mechanisms consistent with the motivations underlying the ``Scientist AI'' paradigm \citep{bengio2025superintelligent}: mandated testing protocols for high-risk applications, domain-specific agency limits calibrated to risk levels, insurance frameworks that price premiums based on measurable agency characteristics, and hard ceilings on agency levels to prevent societal-scale risks.

\section{Background and Related Work}

{\bf Agentic Misalignment.} Research and testing of frontier LLMs have demonstrated that these models are capable of broad manifestations of misalignment. AI systems have demonstrated reward hacking and strategic deception to meet their objectives \citep{pan2024feedback, meinke2025frontier}. Fine-tuning models from narrowly scoped misaligned content has been shown to induce broad misalignment in the model \citep{betley2025emergent, chua2025thought}, and misaligned parent models have passed down misaligned characteristics to their child models \citep{cloud2025subliminal}. With regard to agents, AI systems have also engaged in goal-guarding and sandbagging \citep{meinke2025frontier, vanderweij2024sandbagging}, and have resisted attempts to shut them down using extreme measures \citep{lynch2025agentic}. Misalignment is difficult to detect, is not the result of faulty architecture or training, greatly diminishes the usefulness of an AI system, and is the ordinary result of creating these AI systems through machine learning \citep{dung2023current}. Analysis of misalignment also shows that the risk posed by misaligned systems only increases as these systems become more capable \citep{dung2023current}. This has led to an increasing need to control the level of agency in newly developed LLM-based agents. 

{\bf Scientist AI.} \citet{bengio2025superintelligent} proposed the creation of ``Scientist AI,'' or AI models that are non-agentic, trustworthy, and safe by design. These systems would behave like a scientist, explaining the world from observations but never taking action to please humans or fulfill ideals. The authors state that such systems could aid in scientific research and protect against misaligned AI agents, recognizing that agent development is likely to continue despite the risks. Recognizing this eventuality, we see a growing need to quantify and control the level of agency in an AI system.

{\bf AI Agent Benchmarks Conflate Intelligence and Agency.} AI agent benchmarks typically measure qualities of agents such as reasoning ability, tool selection, and task completion rates, which conflates measures of intelligence and agency. For example, safety-focused benchmarks evaluate agents' adherence to constraints and their resistance to harmful instructions \citep{zhang2025agentsafetybenchevaluatingsafetyllm}, while others measure goal drift and task deviation over extended interactions \citep{arike2025measuring}. Another example, $\tau$-bench \citep{yao2024tau} examines multi-turn interactions and tool usage, and \citet{kwa2025task} evaluate agents' ability to complete increasingly long and complex tasks. Sophisticated real-world benchmarks such as \citet{liu2024agentbench, boisvert2025workarenacompositionalplanningreasoningbased, xu2024theagentcompany} primarily measure task completion rates. Meanwhile, benchmarks that test long-term coherence, such as Vending-Bench \citep{backlund2025vendingbenchbenchmarklongtermcoherence}, reveal how agents fail to maintain consistent autonomous behavior over extended periods, yet still do not provide a framework for measuring agency as a property distinct from intelligence.

{\bf Representation Engineering.} Representation engineering (RepE) is an approach to top‑down transparency that treats representations at the population level, not individual circuits, as the primary object for monitoring and steering abstract concepts and safety-relevant variables (e.g., honesty, harmfulness, power-seeking) \citep{zou2025representationengineeringtopdownapproach}. These techniques have enabled practical applications such as increasing truthfulness \citep{li2024inferencetimeinterventionelicitingtruthful}, reducing sycophancy \citep{papadatos2024linear}, improving instruction-following \citep{stolfo2025improvinginstructionfollowinglanguagemodels}, and allowing the user to control the model's representation of the user \citep{chen2024designingdashboardtransparencycontrol}.

\section{Measurement and Control of the Agency of an LLM-based Agent}
\label{sec:measurement_and_control_of_agency}

Our proposed approach views agency as a measurable and controllable system property distinct from measures of intelligence \citep{hutter2000theoryuniversalartificialintelligence, chollet2019measureintelligence, morris2024levelsagioperationalizingprogress, chollet2024arc}. In Section~\ref{sec:dimensions_of_agency}, we establish a three-dimensional conceptualization of agency that captures the core attributes common to interdisciplinary conceptions of agency while remaining operationalizable through current RepE techniques. In Section~\ref{sec:agency_sliders}, we demonstrate how adapting the approach of \citet{chen2024designingdashboardtransparencycontrol}, can enable both measurement and control of these dimensions through ``agency sliders.'' This white-box approach operates on internal representations rather than outputs, providing robustness against deceptive output-based behaviors while enabling precise calibration of agency levels to match deployment contexts. The resulting framework not only provides the technical foundation for safer LLM-based agent deployment, but also establishes the measurable parameters necessary for the regulatory mechanisms we explore in Section~\ref{sec:regulation_opportunities}.

\subsection{Dimensions of Agency}
\label{sec:dimensions_of_agency}

There are wide-ranging conceptions of agency that span the fields of biology \citep{okasha2024agentbiology, difrisco2025biologicalagency}, philosophy \citep{markus2019thestanford, perezosorio2020intentionalstance, ferrero2022handbookagency}, psychology \citep{moore2016sense}, law \citep{ayres2024law, spamann2024agency}, and computer science \citep{franklin1997agenttaxonomy, kenton2022discoveringagents, chan2023harms, toner2024chat}. For our purposes, we propose an initial set of three dimensions that appear most frequently in the literature. We do not claim that the dimensions are ultimately sufficient, but we offer them as a starting point to operationalize a conception of agency independent of intelligence. These dimensions are: preference rigidity, independent operation, and goal persistence. 

These three dimensions of agency manifest in LLM-based agents as they act in the real-world. Preference rigidity captures how consistently a system maintains preferences and decision patterns across tasks (such as prioritizing certain solution approaches or quality standards regardless of immediate task requirements), independent operation reflects the degree to which a system can proceed without continuous human oversight by making reasonable assumptions from context and available data rather than repeatedly seeking clarification, and goal persistence describes how a system continuously pursues its goals despite obstacles by adapting strategies while maintaining focus on the core goal even when initial approaches fail.

\subsection{Agency Sliders}
\label{sec:agency_sliders}

Building on \citet{chen2024designingdashboardtransparencycontrol}'s representation engineering approach for controlling conversational LLM attributes, we propose developing ``agency sliders'' that enable real-time measurement and control of an LLM-based agent's agentic behavior. \citet{chen2024designingdashboardtransparencycontrol} demonstrated that linear probes trained on residual stream activations can successfully identify and control internal representations of user attributes (age, gender, education, socioeconomic status), allowing users to dynamically adjust these representations through an intuitive dashboard interface. 


Concretely, we propose adapting \citet{chen2024designingdashboardtransparencycontrol} to target our three dimensions of agency where each slider sets a target scalar \(s_d^\star \in [-1,1]\) for dimension \(d \in \{\text{rigidity}, \text{independence}, \text{persistence}\}\). At inference time, a control probe, trained on multi-step scenarios of each dimension, supplies a direction \(v_{d,\ell}\) in the residual stream at selected layers \(\ell\) (chosen by maximizing causal effect size on held-out intervention tasks), and the system applies representation translation \(h_\ell \leftarrow h_\ell + \alpha_d v_{d,\ell}\) with \(\alpha_d\) determined by a closed-loop controller that drives a companion \textit{reader} probe’s score toward \(s_d^\star\). Units are calibrated so that \(|\alpha_d|=1\) corresponds to a one-standard-deviation shift of the reader logit on a calibration suite, resulting in slider semantics that are stable across models and deployments.

This approach offers significant advantages over output-based monitoring or prompt engineering for controlling agent behavior. First, white-box activation-level control is more robust against deceptive behaviors that could compromise output monitoring, as the probes operate on internal representations rather than final outputs that agents might strategically manipulate. Second, representation translation using control probe weight vectors enables granular, continuous adjustment of agency levels, akin to a volume dial rather than an on/off switch, allowing deployment contexts to calibrate agent autonomy precisely to their risk tolerance. For instance, agents deployed in critical infrastructure could operate with low preference rigidity and autonomous operation while maintaining high objective persistence for reliability, whereas research assistants might benefit from higher autonomous operation but lower preference rigidity to remain responsive to user guidance.

\section{New Opportunities for Regulation}
\label{sec:regulation_opportunities}

\subsection{Mandated Testing of High-Risk Agents}

Agency, at any level, is a risk in AI systems. By assessing their level of agency, we can better understand their benefits and potential harms \citep{cihon2025measuringautonomy, bengio2025superintelligent}. Since we cannot assume that these systems are safe without proactive testing \citep{kinniment2024evaluatinglanguagemodelagentsrealistic}, regulators should require pre-deployment testing tied to standardized agency levels, which would enable comparison between AI companies and provide accountability for system failures. Similar to stress tests in finance, or crash tests in automobiles, these evaluations would assess whether, under adversarial conditions, the system remains controllable and within its designated agency limits. Such testing creates a safety baseline, ensuring developers demonstrate compliance before release, as recommended by leading AI safety institutes \citep{UKAISI2024}.

\subsection{Domain-dependent Agency Limits}

By understanding the degree of agency at which misalignment occurs and assessing the inherent risks of different deployment contexts, policymakers can establish agency limits tailored to specific industries or applications \citep{kasirzadeh2025characterizingaiagentsalignment}. Unlike speed limits that can depend on the type and condition of the road, this proportionate oversight of risk should be guided by a uniform policy framework, such as the risk taxonomy of the EU AI Act \citep{eu2024aiact}. Treating a system's degree of agency as a deliberate design choice corresponding to its capabilities and operational environment allows even highly capable systems to be constrained to lower levels of agency when deployed in sensitive contexts \citep{bengio2025internationalscientificreportsafety, feng2025levelsautonomyaiagents}. Without uniform metrics, policymakers and standards bodies tackling broad AI system regulation would be left with only broad categorization that may prove either overly stringent or permissive and difficult to implement. 

\subsection{Agency-based Insurance Frameworks}

A quantified system of agency measurement would also provide infrastructure for the development of insurance frameworks for agentic systems. Strong insurance markets have long been a way of allowing market forces to guide an industry towards safety, as was seen in automobiles when insurers pushed for standards such as airbags \citep{albaum2005safetySells}, and have been suggested as a pathway to regulate AI systems \citep{lior2025insuringAIage, henson2025govtAIinsurance, weil2024insuringEmergingRISKS}. Insurance promises to both incentivize safer design and ensure compensation for victims of AI-related harms \citep{lior2025insuringAIage}. By attaching measurable levels of agency to quantifiable risk profiles, insurers could price premiums based not only on a system's agency level but on the characteristics contributing to a system's agency. High-agency systems operating in sensitive domains would therefore face higher premiums unless they implemented safety mechanisms, incentivizing firms to adopt lower-risk designs. Such an approach would shift the regulatory burden to market forces. Regulators could establish baseline requirements for risk disclosure and require a minimum level of insurance, while insurers would drive compliance through financial incentives.

\subsection{Hard Agency Limits for the Prevention of Societal-Scale Risk}
\label{others}

Agency metrics would enable regulators to set enforceable ceilings on how much agency developers can embed in their AI systems. Similarly to emission standards or nuclear material thresholds, policymakers could establish maximum allowed ``agency levels,'' with mandatory hard stops beyond which further development or deployment is prohibited. Scholars have long argued that even small probabilities of catastrophic harm from runaway AI agents are intolerable and justify strict precautionary limits \citep{bengio2023catastrophic}. In the same spirit with which international governance proposals have suggested treaties imposing global compute-based caps on the training of advanced AI models above agreed-upon thresholds, policymakers could impose specific limits of agency level in AI agents if they pose salient levels of existential risk \citep{miotti2023computecap, raman2025intolerableriskthresholdrecommendations, ramiah2025globalregimecomputegovernance}. By codifying these red lines, policymakers would reduce the likelihood of unauthorized action by agents and the emergence of uncontrollable agentic systems, as in \citet{mitre2025artificial}. It would also provide clear and enforceable limits for AI companies that ensure innovation proceeds within safe and socially acceptable limits.

\section{Conclusion}

In this short paper, we proposed measuring and controlling the agency of LLM-based agents as a system property operationalized through preference rigidity, independent operation, and goal persistence via "agency sliders." This approach enables regulatory mechanisms including mandated testing, domain-specific limits, agency-based insurance, and hard ceilings for social-scale risks. Although our approach requires empirical validation across domains and agent architectures, we believe it represents a significant step toward making agency the direct target of technical intervention and regulatory governance, essential as these agents are increasingly integrated into critical systems.

\bibliographystyle{plainnat}  
\bibliography{references}  

\end{document}